\documentclass{article}
\usepackage{amssymb}

\usepackage{graphicx}
\usepackage{amsmath}

\newtheorem{theorem}{Theorem}

\newtheorem{definition}[theorem]{Definition}
\newtheorem{example}[theorem]{Example}

\newtheorem{notation}[theorem]{Notation}
\newtheorem{problem}[theorem]{Problem}

\newtheorem{remark}[theorem]{Remark}

\newtheorem{summary}[theorem]{Summary}
\newenvironment{proof}[1][Proof]{\textbf{#1.} }{\ \rule{0.5em}{0.5em}}
\input{tcilatex}

\begin{document}

\title{\textbf{QUANTUM PHYSICAL SYSTEMS AS CLASSICAL SYSTEMS \ }}
\author{\textbf{Antonio Cassa} \\
cassa@alpha.science.unitn.it}
\maketitle

\begin{abstract}
A physical system showing a classical (deterministic) behaviour to an
observer can appear to be a quantum system to another observer unable to
distinguish between some distinct states.
\end{abstract}

\begin{summary}
\bigskip Quantum mechanics is a very precise and powerful physical theory
but is accompanied with the negative hypothesis that the measuring process
can have only an essentially statistical, nondeterministic character.

It is hard to believe that in the future this assumption will not be
overcome or reduced in some way by new experiments or new theories: it does
not seem there is any conclusive reason to exclude it.

But is a theory conceivable where the outcomes of the measurements are
uniquely defined and the statistical previsions of quantum mechanics exactly
respected ?

From a mathematical viewpoint it is not difficult to produce an object with
these properties. More difficult would be to justify in physical terms the
artificial construction we propose; however we give a general argument
showing how the interplay between the classical and quantum mechanics we
offer is interpretable as the difference between an imaginary very expert
observer and another nonexpert observer compelled to confuse different
states or different observables.

The main goal of this article is precisely to give a rigorous meaning to
this interplay and a proof that the general quantum system, with all its
states and observables, can be obtained from some classical system.

We cannot offer any physical representation for this classical system, we
confine here proving that besides the well known theorems concerning the
impossibility of hidden variables (cfr. [Neu], [J-P]) there is also room for
a result in favor of the possibility.

All this is made inside the usual descriptions of  the standard quantum
physical systems via quantum logic (cfr. [Mac], [Jau], [Lud] etc.) and,
except for the requirement of hidden variables, does not refer to any
nonhorthodoxical physical theory.
\end{summary}

\bigskip

\section{\textbf{\ Reduction}}

\begin{definition}
\ {A (model for a) \textbf{classical physical system} is a couple $(S,%
\mathcal{L)}$ of a set $S$ (the \textbf{set of pure states}) and a family $%
\mathcal{L}$ of subets of $S$ (the family of \textbf{propositions} of $S$)
distinguishing the elements of $S$ (that is for every couple of different
states there is a proposition in $\mathcal{L}$ not containing both of them).}
\end{definition}

\ 

\ 

Every $L$ in $\mathcal{L}$ represents the subset where a proposition (an
observable taking only the value $0$ and $1$) is true.

\ 

The hypothesis that $\mathcal{L}$ distinguishes the elements of $S$ is made
to simplify the situation; we can always suppose this hypothesis verified
because otherwise we pass to consider as states the classes of the following
equivalence relation: two elements of $S$ are equivalent if every time a
proposition contains one of them it contains them both.

\ 

\begin{example}
Usually $\mathcal{L}$, in the classical case, is the set of all parts of $S$
or the set of all measurable Borel subset of a Borel family. \ 
\end{example}

\bigskip

We are not going to make any restriction on the family $\mathcal{L}$.

\ 

\begin{definition}
\ {An \textbf{observable} for a classical system $(S,\mathcal{L)}$ is a
function $f:S\longrightarrow \mathbf{R}$ such that for every Borel subset $B$
in the Borel family $\mathcal{B}(\mathbf{R})$ of $\mathbf{R}$ the inverse
image $f^{-1}(B)$ is in $\mathcal{L}$.}
\end{definition}

\ 

\begin{example}
In particular the characteristic functions of the subsets $L$ in $\mathcal{L}
$ are observable functions.
\end{example}

\ 

Let's denote by $\mathcal{F}$ the set of all the observable functions of the
system $(S,\mathcal{L)}$.

\ 

\begin{remark}
If $f$ is an observable function of $(S,\mathcal{L)}$ and $g:\mathbf{R}%
\longrightarrow \mathbf{R}$ is a Borel function the function $g\circ
f:S\longrightarrow \mathbf{R}$ is an observable. Infact for every Borel
subset $B$ the set $(g\circ f)^{-1}(B)=f^{-1}(g^{-1}(B))$ is in $\mathcal{L}$%
.
\end{remark}

\ 

\ 

These definitions are given to idealize the situation of an observer (that
we will call the {\normalsize \ }\emph{precise observer}) able to prepare
with extreme precision a physical system in a variety of different (pure)
states and to perform on the system several measures in such a way that when
he prepares the state $s$ and performs the observable $f$ he can get the
exact value $f(s)$. Our observer knows all the time and exactly what state
is preparating and what observable is performing.

\ 

To have a more precise idea of this kind of situation let us suppose that
the precise observer has a very huge and efficient laboratory where he can
prepare every sort of state of the physical system under consideration and
use every sort of measuring apparatus: all he has to do is to give to the
laboratory's computer a ''string'' specifying completely and exactly the
state to prepare and another ''string'' specifying the observable to measure
and the fantastic laboratory does all the work. \ 

The observer checks that given a ''state string'' and an ''observable
string'' he always gets the same value and so he can state with certainty
that the physical system considered is \emph{classical }(deterministic).

\ 

Let us consider now another observer (that we will call the \textbf{\ }\emph{%
imprecise observer}) studying the same physical system but with a poorer
ability; this second observer can produce all the states and observables of
the previous one but he does not know exactly what he makes: he gives a
''procedure'' to produce a certain state and another ''procedure'' to
produce the measuring apparatus but if he repeats the given procedures he
can get different values in a random and, for him, unavoidable way.

\ 

Let us suppose moreover that the precise observer can describe precisely
what the problem is with the imprecise observer: when this second one
chooses a ''procedure'' he produces a state among several different ones in
a given class of equivalence of $S$ with a certain probability: there is an
equivalence relation (the \textbf{confusion} relation) $\mathcal{R}$ on the
classical system $(S,\mathcal{L)}$ and a probability measure $\mu _{p}$ on
every equivalence class $p$ in the quotient set $P=S/\mathcal{R}$. When the
imprecise observer tries to prepare the system with a given procedure $p$ he
does not know which one of the states in the class $p=[s]$ he is really
preparing, therefore when he evaluates the observable $f$ he can get any one
of the values in the subset $f([s])$. Making several trials he experiments
all these values with different frequencies arriving in the end at the
conclusion that the measure of the observable $f$ on the ''preparation'' $p$
has a statistical character and that he cannot get anything more that the
probability $\pi (f,p,B)$ that the measure of $f$ on $p$ lies in the Borel
subset $B$ of $\mathbf{R}$.

\ 

For the precise observer it is obvious that 
\begin{equation*}
\pi (f,p,B)=\mu _{p}(f^{-1}(B)\cap p)
\end{equation*}
.

\ 

If the imprecise observer is left unaware of his ''confusion'' and convinced
that he cannot get any more information on the system, he will decide,
coherently, not to distinguish between preparations or measuring apparatuses
giving the same probabilities. Therefore he will define the following
concept:

\bigskip

\begin{definition}
\ A (model for a) \textbf{statistical physical system} is a triple $(P,%
\mathcal{O,\pi )}$ of a set $P$ (the set of statisical \textbf{states}),
another set $\mathcal{O}$ (the set of statistical \textbf{observables}) and
a function: $\pi :\mathcal{O\times P\times B(\mathbf{R})\longrightarrow }$ $%
[0,1]$ (the \textbf{probability} that the measure of a observable on a state
lies in a Borel subset of $\mathbf{R}$) such that

\begin{enumerate}
\item  $\pi (T^{\prime },p,B)=\pi (T^{\prime \prime },p,B)$ for every $p$ in 
$P$ and $B$ in $\mathcal{B(\mathbf{R})}$ implies $T^{\prime }=T^{\prime
\prime }$ and

\item  $\pi (T,p^{\prime },B)=\pi (T,p^{\prime \prime },B)$ for every $T$ in 
$\mathcal{O}$ and $B$ in $\mathcal{B(\mathbf{R})}$ implies $p^{\prime
}=p^{\prime \prime }$.
\end{enumerate}
\end{definition}

\ 

From the viewpoint of the precise observer this means that two states $%
s^{\prime }$ and $s^{\prime \prime }$ are equivalent in the equivalence
relation of confusion if and only if 
\begin{equation*}
\mu _{\lbrack s^{\prime }]}(f^{-1}(B)\cap \lbrack s^{\prime }])=\mu
_{\lbrack s^{\prime \prime }]}(f^{-1}(B)\cap \lbrack s^{\prime \prime }])
\end{equation*}
for every $f$ in $\mathcal{F}$ and every $B$ in $\mathcal{B(\mathbf{R})}$. \ 

Moreover the precise observer makes a discovery:  the imprecise observer
confuses not only the states but also the observables. \ 

The set of statistical observables $\mathcal{O}$ is the quotient set of $%
\mathcal{F}$ modulo the equivalence relation stating that two functions $%
f^{\prime }$ and $f^{\prime \prime }$ of $\mathcal{F}$ are equivalent if 
\begin{equation*}
\mu _{\lbrack s]}(f^{\prime -1}(B)\cap \lbrack s])=\mu _{\lbrack
s]}(f^{\prime \prime -1}(B)\cap \lbrack s])
\end{equation*}
for every $s$ in $S$ and every $B$ in $\mathcal{B(\mathbf{R})}$.

\ 

Therefore the precise observer can give the following:

\ 

\begin{definition}
\ A \textbf{confusion} relation for a classical system $(S,\mathcal{L)}$ is
given assigning an equivalence relation $\mathcal{R}$ on $S$ and a
probability measure $\mu _{p}$ on every equivalence class in such a way that
for every couple of inequivalent elements $s^{\prime }$ and $s^{\prime
\prime }$ in $S$ there exists a proposition $L$ in $\mathcal{L}$ with $\mu
_{\lbrack s^{\prime }]}(L\cap \lbrack s^{\prime }])\neq \mu _{\lbrack
s^{\prime \prime }]}(L\cap \lbrack s^{\prime \prime }])$.
\end{definition}

\ 

\ 

It is clear that for every confusion relation $\mathcal{R}$ there is also
defined  an equivalence relation $\mathcal{M}$ on the set $\mathcal{F}$ of
observable functions by taking $f^{\prime }\mathcal{M}f^{\prime \prime }$
if: 
\begin{equation*}
\mu _{\lbrack s]}(f^{^{\prime }-1}(B)\cap \lbrack s])=\mu _{\lbrack
s]}(f^{^{^{\prime \prime }-1}}(B)\cap \lbrack s])
\end{equation*}
for every $s$ in $S$ and every $B$ in $\mathcal{B(\mathbf{R})}$. \ 

Therefore a statistical system is well defined by taking

\begin{itemize}
\item[1)]  $\widehat{S}=S/\mathcal{R}$

\item[2)]  $\widehat{\mathcal{F}}=\mathcal{F}/\mathcal{M}$

\item[3)]  $\widehat{\mu }:\widehat{\mathcal{F}}\times \widehat{S}\times 
\mathcal{B(\mathbf{R})\longrightarrow }$ $[0,1]$ given by $\widehat{\mu }%
([f],[s],B)=\mu _{\lbrack s]}(f^{-1}(B)\cap \lbrack s])$.
\end{itemize}

\ 

\begin{definition}
\ Given a classical system $(S,\mathcal{L)}$ and a confusion relation

$(\mathcal{R},\{\mu _{p}\}_{p\in S/\mathcal{R}})$, the statistical system $(%
\mathcal{F}/\mathcal{M},\mathcal{S}/\mathcal{R},\widehat{\mu })$ is called
the \textbf{system reduced } by the confusion relation.
\end{definition}

\ 

\begin{remark}
We call the procedure given above the reduction  just because we pass from a
state space to another making a quotient along (essentially) one-dimensional
fibres as when we reduce a contact manifold producing a symplectic manifold.
\end{remark}

\ 

When a statistical system can be obtained as a reduced system of a classical
system there is at least a mathematical reason to talk of \textbf{hidden
variables} (the ''variables'' describing the elements in each equivalence
classes of the state set of the classical system): under every statistical
state $p=[s]$ are ''hidden'' the elements of $[s]$, the ''true states''. \ 

It is possible to make precise this assertion considering the following
(cfr. [Jam] pag.262):

\ 

\begin{definition}
\ \textbf{\ }Let $(\mathcal{O},P,\pi )$ be a statistical system, a \textbf{%
model for a system with hidden variables with respect to $(\mathcal{O},P,\pi
)$ } is given assigning:

\begin{enumerate}
\item  a set $S$ (the \textbf{space of hidden states}) a surjective map $%
\rho :S\longrightarrow P$ (associating to a ''hidden state'' its ''apparent
state'')

\item  for each ''apparent'' state $p\in P$ a probability measure $\mu _{p}$
on $S$ (representing the probability to find in a measurable subset of $S$ a
''hidden state'' representing $p$); and

\item  for each observable $T\in \mathcal{O}$ a function $%
f_{T}:S\longrightarrow \mathbf{R}$ (representing a classical observable
giving the values that appears randomly for the statistical observable $T$)
such that for every Borel subset $B$ of $\mathbf{R}$: 
\begin{equation*}
\pi (T,p,B)=\mu _{p}(f_{T}^{-1}(B))
\end{equation*}
(that is the probability that the value of $T$ on $p$ lies in $B$ is given
by the probability to find a hidden state of $p$ between the states where
the observable $f_{T}$ takes a value in $B).$
\end{enumerate}
\end{definition}

\ 

\ 

In fact, in the case of a reduced system, we can take as $\rho: S
\longrightarrow P$ the quotient map and for every $p$ in $P$ as probability
measure $\mu_p$ the measure $\mu_p$ seen as a measure on all $S$ and not
only on $\rho^{-1}(p)$.

\ 

We are going to prove that the general quantum system (given by \ a Hilbert
space) is a reduced system of a classical system. \ 

This kind of property is sometimes considered impossible to be proved or in
contradiction with the principles of the standard quantum mechanics. \ 

On the contrary the same property can be considered ''well known'' and quite
obvious: if you want a ''hidden variable '' function giving the right
statistical outcomes for a self-adjoint operator $T$ of the Hilbert space $%
\mathbf{H}$ simply take the ''quasi-inverse'' function $f:\mathbf{P}(\mathbf{%
H})\times ]0,1[\longrightarrow \mathbf{R}$ (cfr. the proof of the following
theorem.) defined by 
\begin{equation*}
f([h],t)=sup\{u:\langle E_{]-\infty ,u]}^{T}\rangle _{h}\geq t\}.
\end{equation*}
. \ 

It seems that all depends on what you mean. In this section we have just
tried to suggest a plausible interpretation that can save determinism in
observations.

\bigskip

\bigskip

\ 

\section{\ \textbf{The quantum system as a reduced system}}

\ 

\begin{definition}
\ The (model) for the (irreducible) \textbf{quantum system} is given
assigning:

\begin{enumerate}
\item  the (complex) projective space $\mathbf{P}(\mathbf{H})$ of a Hilbert
space $\mathbf{H}$ (of dimension at least two) as state space;

\item  the set $SA(\mathbf{H})$ of self-adjoint operators on $\mathbf{H}$ as
observable space; and

\item  the function $\pi :SA(\mathbf{H})\times \mathbf{P}(\mathbf{H})\times 
\mathcal{B(\mathbf{R})}\longrightarrow \lbrack 0,1]$ defined by 
\begin{equation*}
\mathbf{\pi }(T,[h],B)=\langle E_{B}^{T}\rangle _{h}={\frac{{\langle
h,E_{B}^{T}(h)\rangle }}{{\langle h,h\rangle }}}
\end{equation*}
(where $E_{B}^{T}=\chi _{B}\circ T$ is the projector operator associated to
the Borel subset $B$ of $\mathbf{R}$ in the spectral measure of $T$) as
probability function.\ 
\end{enumerate}
\end{definition}

\ 

\bigskip 

\begin{theorem}
\textbf{\ } The quantum system is the reduced system of a classical system.
\end{theorem}

\ 

\begin{proof}
For every $[h]$ in $\mathbf{P}(\mathbf{H})$ let us consider a complete
separable metric space $S_{[h]}$ with a Borel measure $\mu _{\lbrack h]}$
such that $\mu _{\lbrack h]}(S_{[h]})=1$ and $\mu _{\lbrack h]}(\{s\})=0$
for every $s$ in $S_{[h]}$. For every such space there is a measurable map $%
\phi _{\lbrack h]}:S_{[h]}\longrightarrow ]0,1[$ such that $\phi _{\lbrack
h]\ast }(\mu _{\lbrack h]})=\lambda $ where $\lambda $ denotes the Lebesgue
measure on the interval (cfr. [Roy] Thm. 9 pag. 327).

Let $S$ be the disjoint union of the $\{S_{[h]}\}_{[h]\in \mathbf{P}(\mathbf{%
H})}$. We will call a subset $L$ of $S$ a proposition if $L\cap S_{[h]}$ is
a measurable set for every $[h]$ in $\mathbf{P}(\mathbf{H})$ and if there
exists a projector $E$ of $\mathbf{H}$ such that $\mu _{\lbrack h]}(L\cap
S_{[h]})=\langle E\rangle _{h}$ for every $[h]$ in $\mathbf{P}(\mathbf{H})$.
\ 

Let $\mathcal{L}$ be the set of all propositions in $S$. \ 

Every proposition $L$ determines the set of all $[h]$ where $\mu _{\lbrack
h]}(L\cap S_{[h]})=1$ and therefore the projector $E$. If we denote by $%
\mathcal{E}$ the set of all projectors of $\mathbf{H}$, a map $\varepsilon :%
\mathcal{L}\longrightarrow \mathcal{E}$ associating to a proposition its
projector is well defined . \ 

The map $\varepsilon $ is surjective; fixed $E$ is enough to take in $S_{[h]}
$ a measurable subset $L_{[h]}$ such that $\mu _{\lbrack h]}(L_{[h]}\cap
S_{[h]})=\langle E\rangle _{h}$ and then take $L$ as the disjoint union of
all the $L_{[h]}$. It is not difficult to prove, in a similar way, that the
propositions distinguish the elements in $S$. \ 

Let us denote by $\mathcal{F}$, as usual, the set of all observable
functions for $(S,\mathcal{L})$; we want to prove that to each of these
functions $f$ is associated a self-adjoint operator $T$ such that 
\begin{equation*}
\mu _{\lbrack h]}(f^{-1}(B)\cap S_{[h]})=\langle E_{B}^{T}\rangle _{h}
\end{equation*}
for every $[h]$ in $\mathbf{P(H)}$ and $B$ in $\mathcal{B}(\mathbf{R})$. \ 

For every real number $t$ the proposition $L_{t}=f^{-1}(]-\infty ,t])$
determines a projector $E_{t}$. The family $\{E_{t}\}_{t\in \mathbf{R}}$ is
a spectral family of projectors of $\mathbf{H}$ (cfr. [Wei] def. (7.11) pag.
180); in fact $\langle E_{t}\rangle _{h}=$ $=(f|_{S_{[h]\ast }}\mu _{\lbrack
h]})(]-\infty ,t])$ for every $h$ and therefore the monotonicity, the
left-continuity and the convergence to $0$ and $1$ properties for the
projection operators follow from the analogous properties of cumulative
distribution functions for Borel probabilitity measures (cfr. [Roy] Lemma 10
pag. 262). \ 

Hence the spectral family $\{E_{t}\}_{t\in \mathbf{R}}$ defines a
self-adjoint operator $T$ such that for every $t$ in $\mathbf{R}$ 
\begin{equation*}
\mu _{\lbrack h]}(f^{-1}(]-\infty ,t])\cap S_{[h]})=\langle E_{]-\infty
,t]}^{T}\rangle _{h}
\end{equation*}
and therefore for every Borel subset $B$ of $\mathbf{R}$ 
\begin{equation*}
\mu _{\lbrack h]}(f^{-1}(B)\cap S_{[h]})=\langle E_{B}^{T}\rangle _{h}.
\end{equation*}
The operator $T$ is unambiguously defined by the function $f$, let us denote
by $\tau :\mathcal{F}\longrightarrow SA(\mathbf{H})$ the map so defined. Let
us prove this map is surjective. \ 

For every $[h]$ let us denote by $F_{[h]}:\mathbf{R}\longrightarrow ]0,1[$
the distribution function $F_{[h]}(u)=\langle E_{]-\infty ,u]}^{T}\rangle
_{h}$; its induced Borel measure $\nu _{F_{[h]}}$ has the property that $\nu
_{F_{[h]}}(B)=\langle E_{B}^{T}\rangle _{h}$ for every Borel subset $B$. \ 

Its quasi-inverse $\widetilde{F_{[h]}}:]0,1[\longrightarrow \mathbf{R}$
verifies $\widetilde{F_{[h]}}_{\ast }(\lambda )=\nu _{F_{[h]}}$ (cfr. [K-S]
Thm. 4 pag. 94) and therefore $(\widetilde{F_{[h]}}\circ \phi _{\lbrack
h]})_{\ast }(\mu _{\lbrack h]})=\nu _{F_{[h]}}$, that is 
\begin{equation*}
(\widetilde{F_{[h]}}\circ \phi _{\lbrack h]})_{\ast }(\mu _{\lbrack
h]})(]a,b])=F_{[h]}(b)-F_{[h]}(a)=\langle E_{]a,b]}^{T}\rangle _{h}
\end{equation*}
for every $a<b$ in \textbf{R}. \ 

The function $f:S\longrightarrow \mathbf{R}$ defined by $f(s)=\widetilde{%
F_{[h]}}(\phi _{\lbrack h]}(s))$ (where $[h]$ contains $s$) has the desired
property: $\tau (f)=T$. \ 

Let us prove that the reduced system of $(S,\mathcal{L})$ is the quantum
system. Two elements $r$ in $[h]$ and $s$ in $[k]$ are equivalent if and
only if $\mu _{\lbrack h]}(L\cap S_{[h]})=\mu _{\lbrack k]}(L\cap S_{[k]})$
for every proposition $L$, therefore if and only if $\langle E\rangle
_{h}=\langle E\rangle _{k}$ for every projector $E$ of $\mathbf{H}$, that is
if and only if $[h]=[k]$. \ 

Two functions $f$ and $g$ in $\mathcal{L}$ are equivalent if and only if $%
\langle E_{B}^{\tau (f)}\rangle _{h}=\langle E_{B}^{\tau (g)}\rangle _{h}$
for every $h$ and $B$. This means $E_{B}^{\tau (f)}=E_{B}^{\tau (g)}$ for
every $B$, that is $\tau (f)=\tau (g)$. \ In the end $\widehat{\mu }%
([f],[h],B)=\mu _{\lbrack h]}(f^{-1}(B)\cap S_{[h]})=\langle E_{B}^{\tau
(f)}\rangle _{h}=\pi (\tau (f),[h],B)$.
\end{proof}

\ 

\ 

From now on we will denote by $(S,\mathcal{L)}$ a classical system giving
the (irreducible) quantum system as reduction, by $\mathcal{F}$ its set of
observable functions and by $\rho :S\longrightarrow \mathbf{P}(\mathbf{H})$, 
$\tau :\mathcal{F}\longrightarrow SA(\mathbf{H})$ and $\varepsilon :\mathcal{%
L}\longrightarrow \mathcal{E}$ the quotient maps.

\ 

\begin{remark}
If $L$ is in $\mathcal{L}$ then also $(S\setminus L)$ is in $\mathcal{L}$
and $\varepsilon (S\setminus L)=I-\varepsilon (L)$.
\end{remark}

\ 

\ 

\begin{remark}
If $T$ is a self-adjoint operator with spectral measure $\{B\mapsto E_{B}\}$
and $g:\mathbf{R}\longrightarrow \mathbf{R}$ is a Borel function then a
self-adjoint operator $g(T)$ with spectral measure $\{B\mapsto
E_{g^{-1}(B)}\}$ is well defined .
\end{remark}

\ 

\ 

\begin{theorem}
If $f$ is an observable function of a classical system $(S,\mathcal{L)}$
reducing to the quantum system and $g:\mathbf{R}\longrightarrow \mathbf{R}$
is any Borel function, it holds $\tau (g\circ f)=g(\tau (f))$.
\end{theorem}

\ 

\begin{proof}
The function $f$ gives the operator $\tau (f)=T$ with spectral measure%
\newline
$\{B\mapsto E_{B}=\varepsilon (f^{-1}(B))\}$. The observable function $%
g\circ f$ defines the spectral measure $\{B\mapsto \varepsilon
(f^{-1}(g^{-1}(B)))\}$ and this is exactly the spectral measure of $g(T)$.
\end{proof}

\ 

\ 

\begin{theorem}
If $f$ is an observable function of a classical system $(S,\mathcal{L)}$
reducing to the quantum system and $g:\mathbf{R}\longrightarrow \mathbf{R}$
is any Borel function, it holds

\begin{enumerate}
\item  $\langle g(\tau (f))\rangle _{h}=\int_{0}^{1}g(f([h],t))\cdot d\
\lambda (t)$

\item  $\langle \tau (f)\rangle _{h}=\int_{0}^{1}f([h],t)\cdot d\ \lambda (t)
$
\end{enumerate}
\end{theorem}

\ 

\begin{proof}
The point (2) follows from (1) taking $g=id_{\mathbf{R}}$. Let us prove (1): 
\begin{equation*}
\langle g(\tau (f))\rangle _{h}=\int_{\mathbf{R}}g(u)\cdot d\ \nu _{\langle
E_{]-\infty ,.]}^{\tau (f)}\rangle _{h}}=\int_{\mathbf{R}}g(u)\cdot (d\
f_{[h]\ast }\lambda _{]0,1[})=\int_{]0,1[}g(u)\circ f_{[h]}\cdot d\ \lambda .
\end{equation*}
Cfr. [Wei] Thm. 7.14(e) and [K-S] Cor. 3 pag. 93 for the passages.
\end{proof}

\ 

\ 

\ 

\begin{theorem}
Given two projectors $E$ and $F$, if there exist two propositions $L$ and $M$
in $\mathcal{L}$ such that

\begin{enumerate}
\item  the (finite) boolean algebra of subsets $\mathcal{A}$ generated by $L$
and $M$ is contained in $\mathcal{L}$ and

\item  the map $\varepsilon :\mathcal{A\longrightarrow E}$ sends $L$ in $E$, 
$M$ in $F$ transforming the operation $\wedge $ in $\cap $, $\vee $ in $\cup 
$ and the complementation in the orthogonality

then the projectors $E$ and $F$ are compatible (that is commute).
\end{enumerate}
\end{theorem}

\ 

\begin{proof}
The union $L\cup M=L\cup (\ \complement L\cap M)=M\cup (\mathcal{\complement 
}M\cap L)$ belongs to $L$ and 
\begin{eqnarray*}
\varepsilon (L\cup M) &=&E\vee F=\varepsilon (L\cup (\complement L\cap
M))=E\vee (E^{\prime }\wedge F)= \\
&=&\varepsilon (M\cup (\mathcal{\complement }M\cap L))=F\vee (F^{\prime
}\wedge E).
\end{eqnarray*}
This proves that $E$ and $F$ are compatible (cfr. [Jau] probl.2 of 5-8, pag.
87).
\end{proof}

\ 

\begin{remark}
Therefore whenever you consider two noncommuting projectors $E$ and $F$ it
is impossible to find two propositions $L$ and $M$\bigskip\ with the
properties (1) and (2) of the previous theorem.
\end{remark}

.

\begin{definition}
\ Let $E_{1},E_{2}$ and $F_{1},F_{2}$ be two couples of projectors in $%
\mathcal{E}$. We will say that the couples \textbf{admit proposition
intersections} if there are two couples of propositions $A_{1},A_{2}$ and $%
B_{1},B_{2}$ with $\varepsilon (A_{i})=E_{i}$ and $\varepsilon (B_{j})=F_{j}$
for $i,j=1,2$ and such that the 16 intersections $A_{i}\cap B_{j}$, $%
\complement A_{i}\cap B_{j}$, $A_{i}\cap \complement B_{j}$, $\complement
A_{i}\cap \complement B_{j}$ are all in $\mathcal{L}$ and $\varepsilon
(A_{i}\cap B_{j})=E_{i}\wedge F_{j}$ , $\varepsilon (\complement A_{i}\cap
B_{j})=(I-E_{i})\wedge F_{j}$, $\varepsilon (A_{i}\cap \complement
B_{j})=E_{i}\wedge (I-F_{j})$ and $\varepsilon (\complement A_{i}\cap
\complement B_{j})=(I-E_{i})\wedge (I-F_{j})$.
\end{definition}

\begin{notation}
In this situation we will consider the self-adjoint operators $T_{ij}(E,F)=%
\newline
=E_{i}\wedge F_{j}+(I-E_{i})\wedge (I-F_{j})-(I-E_{i})\wedge
F_{j}-E_{i}\wedge (I-F_{j})=\newline
=\varepsilon (A_{i}\cap B_{j})+\varepsilon (\complement A_{i}\cap
\complement B_{j})-\varepsilon (\complement A_{i}\cap B_{j})-\varepsilon
(A_{i}\cap \complement B_{j})$.
\end{notation}

\ 

\ 

\begin{theorem}
If $E_{1},E_{2}$ and $F_{1},F_{2}$ are two couples of projectors in $%
\mathcal{E}$ admitting proposition intersections, then for every $h$ in $%
\mathbf{H}\setminus \{0\}$ it holds the inequality 
\begin{equation*}
|\langle T_{11}(E,F)\rangle _{h}-\langle T_{12}(E,F)\rangle _{h}|+|\langle
T_{21}(E,F)\rangle _{h}+\langle T_{22}(E,F)\rangle _{h}|\leq 2.
\end{equation*}
\ 
\end{theorem}

\begin{proof}
The functions: $f_{ij}=\chi _{A_{i}\cap B_{j}}+\chi _{(S\setminus A_{i})\cap
(S\setminus B_{j})}-\chi _{(S\setminus A_{i})\cap B_{j}}-\chi _{A_{i}\cap
(S\setminus B_{j})}$ are functions on $S$ with $\int_{0}^{1}f_{ij}([h],t)%
\cdot d\ \lambda (t)=\langle T_{ij}(E,F)\rangle _{h}$. \ It is not difficult
to check in $S$ the following equality: $|f_{11}-f_{12}|+|f_{21}+f_{22}|=2.$
\ 

Therefore,

$|\langle T_{11}(E,F)\rangle _{h}-\langle T_{12}(E,F)\rangle _{h}|+|\langle
T_{21}(E,F)\rangle _{h}+\langle T_{22}(E,F)\rangle _{h}|=\newline
=|\int_{0}^{1}f_{11}([h],t)\cdot d\ \lambda
(t)-\int_{0}^{1}f_{12}([h],t)\cdot d\ \lambda
(t)|+|\int_{0}^{1}f_{21}([h],t)\cdot d\ \lambda (t)+$

$+\int_{0}^{1}f_{22}([h],t)\cdot d\ \lambda (t)|\leq
\int_{0}^{1}(|f_{11}-f_{12}|+|f_{21}+f_{22}|)([h],t)\cdot d\ \lambda (t)=2$. 
\textbf{.}
\end{proof}

\ 

\ 

\begin{remark}
The proof of the previous theorem mimics the usual one given to prove one of
the Bell inequalities and in fact if you take in a quantum system two
couples of projections not verifying the Bell inequality for some state, you
have two couples of projections not admitting proposition intersections. \ 
\end{remark}

\ 

\begin{problem}
When we have two couples of projections not admitting proposition
intersections we can consider the system as the reduction of a classical
system and replace the projectors $E_{1},E_{2},F_{1},F_{2}$ with some
propositions $A_{1},A_{2},B_{1},B_{2}$, but we do not dispose of, for
example, the proposition $A_{1}\cap B_{1}$ (or this intersection does not
correspond to the projector $E_{1}\wedge F_{1}$). \ 

The absence of $A_{1}\cap B_{1}$is in contrast with the possibility,
considered natural in a classical physical theory, to check ''in the same
time'' two properties of a system.

This is undoubtly strange and uncomfortable, however, if we take seriously
the hypothesis of the precise observer, any objection to this eventuality
cannot be considered definitive unless expressed in terms of his physics: in
other words, we should be able first to know his description of the physical
reality and how he can explain, for example, a possible absence of
intersections.
\end{problem}

\ 

\bigskip

\textbf{BIBLIOGRAPHY}

\ 

\ 

[C] I. Calvino: The nonexistent knight. \ 

Harcourt Trade Publishers, London 1988

\ 

[Jam] M. Jammer: The philosophy of quantum mechanics. \ 

John Wiley \& sons, NY 1974

\ 

[Jau] J. M. Jauch: Foundations of quantum mechanics. \ 

Addison-Wesley P.C., Reading (Mass.) 1968

\ 

[J-P] J. M. Jauch and C. Piron: Can hidden variables be excluded in quantum
mechanics?

Helv. Phys. Acta 36, 827-837 (1963)

\ 

[K-S] J. K. Kelley and T.P. Srinivasan: Measure and integral. \ 

Springer-Verlag, NY 1988

\ 

[Lud] G. Ludwig: Foundations of quantum mechanics, vol. I

Springer-Verlag, NY 1983

\ 

[Mac] G. Mackey: Mathematical foundations of quantum mechanics

Benjamin, NY 1963

\ 

[Neu] J. von Neumann: Mathematical foundations of quantum mechanics.

Princeton univ. press, Princeton NJ 1955

\ 

[Pir] C. Piron: Foundations of quantum physics.

Mathematical Physics monograph series (ed. A Wightman).

Benjamin, Reading 1976

\ 

[Roy] H.L. Royden: Real analysis. \ 

The Macmillan co., London 1968

\ 

[Wei] J. Weidmann: Linear operators in Hilbert spaces. \ 

Springer-Verlag, NY 1980

\ 

\ 

\end{document}